\documentclass[11pt]{amsart}
\usepackage{anyfontsize}
\usepackage{tikz-cd}
\usepackage[headings]{fullpage}
\usepackage{amsmath}
\usepackage{amsfonts}
\usepackage{latexsym}
\usepackage{amssymb}
\usepackage{placeins}

\newtheorem{proposition}{Proposition}[section]
\newtheorem{theorem}{Theorem}[section]

\newtheorem{lemma}{Lemma}[section]
\newtheorem{definition}{Definition}[section]
\newtheorem{corollary}{Corollary}[section]

\def\ker{\mathop{\sf Ker}}

\def\im{\mathop{\sf Im}}
\def\coker{\mathop{\sf Coker}}

\def\dim{\mathop{\sf dim}}

\def\B ellk{\mathcal{B}_{\ell,k}}

\def\C{\mathbb{C}}

\def\Z{\mathbb{Z}}

\def\aut{\mathop{\sf Aut}}
\def\gl{\mathop{\sf GL_n(\mathbb{Z})}}

\def\trace{\mathop{\sf Tr_\alpha}}
\def\tracep{\mathop{\sf Tr'_\alpha}}
\def\tracei{\mathop{\sf Tr_{\alpha_i}}}
\def\traceo{\mathop{\sf Tr_{\alpha_1}}}
\def\tracet{\mathop{\sf Tr_{\alpha_2}}}
\def\log{\mathop{\sf log}}

\newcommand{\F}{\mathcal{F}}

\begin{document}

\title [ISP for the Laurent ring]{The Indefinite Summation Problem for the Laurent ring} 

\author
[S. Shankar]{Shiva Shankar} 
\address{Kerala School of Mathematics, Kozhikode, India}
\email{shunyashankar@gmail.com }

\begin{abstract} This article solves the Indefinite Summation Problem (ISP) for the difference ring $(A, \alpha)$, where $A$ is the Laurent ring of shift operators on the lattice $\Z^n$, and $\alpha$ is any ring automorphism of $A$ of finite order. The space of summable elements in $(A, \alpha)$ is infinite dimensional, and it is possible to write down an explicit basis for it. The  solution translates to a finite procedure involving a matrix multiplication, where the size of the matrix can be estimated. It follows that the arithmetic complexity of the solution can also be determined. These results extend to a solution of the ISP for the ring of functions on $\Z^n$, on which $\alpha$ acts by duality.

The article also points out that the solution to the ISP amounts to calculating the group cohomologies $H^i([\alpha], A), i = 0, 1$, where $[\alpha]$ is the cyclic group generated by $\alpha$. 

\vspace{2.5mm} 
\noindent {\tiny Key words: Indefinite summation, Computational algebra, Symbolic computation, Laurent polynomial ring}

\vspace{1.5mm}
\noindent {\tiny AMS Subject Classifications: 68W30, 33F10, 16S34}

\end{abstract}

\maketitle

\section{introduction}

A central question in Symbolic Computation is the Indefinite Summation Problem (ISP). Its origins lie  in the discrete analogue of the indefinite integral, and asks when a function $g: \Z \rightarrow \C$ equals the forward difference $\Delta(f)$ of some $f: \Z \rightarrow \C$, where $\Delta(f)(x) = f(x+1) - f(x)$. With rapid advancements in computing machinery, and the consequent developments in algorithms and in computer algebra, this problem evolved to questions in symbolic computation involving the existance of closed form solutions, questions which are discrete versions of questions in symbolic integration.  

Usually, the problem also specifies the class of functions to which $g$ belongs, and requires $f$ to be from the same class, or in some given extension of it. For instance, \cite{a} provides an algorithmic solution to the ISP for the class of rational functions. A decision method for the case of hypergeometric summation, when it is required that $\frac{f(x+1)}{f(x)}$ be rational, is the work in \cite{go}. These results have been significantly generalised by Chyzak to the case of holonomic sequences in \cite{chy}.
\\

This article is in the setting of an `abstract' version of the ISP due to Chen et al. \cite{chen}. To state it, we rewrite the classical ISP in the following notation: let $\F$ be the ring of functions on $\Z$, and let $\alpha$ be the automorphism of $\F$ given by the shift operator $\alpha(f)(x) = f(x+1)$. Then, the ISP asks when a function $g$ is in the image of $\alpha - 1$.

\vspace{1mm}

In this notation, the abstract ISP can be stated thus \cite{chen}:

 \vspace{1mm}
 
Let $A$ be a ring, and $\alpha: A \rightarrow A$ an automorphism of $A$. The pair $(A, \alpha)$ is called a difference ring. The constant subring of $(A,  \alpha)$ is the set $\{a\in A~|~\alpha(a)=a\}$ of fixed points of $\alpha$. A difference ring $(A^*, \alpha^*)$ is a difference extension of $(A, \alpha)$ if $A \subset A^*$ and $\alpha^*|_A = \alpha$ (the extended automorphism $\alpha^*$ of $A^*$ is also denoted by $\alpha$). 
 \vspace{.8mm}
 
{\em The (Abstract) ISP}: Let $(A^*, \alpha)$ be a 
 specific difference extension of $(A, \alpha)$. 
 Given $b \in A$, decide whether there exists $a \in A^*$ such that $\alpha(a) - a = b$. 
 If such an $a$ exists, $b$ is said to be summable in $A^*$.\\
 
 This article provides a solution to the ISP in this setting, when $A$ is the Laurent ring generated by the forward and backward coordinate shifts of the lattice $\Z^n$, and 
 where $\alpha$ is any automorphism of $A$ of finite order. It includes both a decision method to determine when an element $b \in A$ is summable, as well as a finite proceedure to construct solutions $a$ to $\alpha(a) - a = b$ when $b$ is summable. 
 
 The article is organised as follows: in the next section, we consider the finite cyclic group $[\alpha]$  generated by $\alpha$, and describe the structure of the orbits of its action on $A$. We can now calculate the trace of an element of $A$ with respect to $\alpha$, and then establish a necessary and sufficient condition for the solution to the ISP as in Section 3. The space of summable elements is infinite dimensional, and we write down an explicit basis for it.
This calculation also shows that $H^1([\alpha], ~A)$, the first group cohomology of $[\alpha]$ with values in $A$, vanishes. 

Given a summable element $b$, and one solution $a_1$ to $\alpha(a) - a = b$, every other solution is obtained by adding a constant to $a_1$. These constants are also the fixed points of the action of $[\alpha]$ on $A$. We describe the set of these fixed points in Section 4; it is the zero-th group cohomology $H^0([\alpha], A)$. Thus, the solution to the ISP for $(A, \alpha)$ is also the calculation of the low dimensional group cohomologies of $[\alpha]$ with coefficients in the $[\alpha]$-module $A$.
 
As the automorphism $\alpha$ is of finite order, and as an element in $A$ has a finite number of monomial terms, there is a finite decision method, involving elementary matrix calculations, to determine whether an element of $A$ is summable. This decision method, together with its arithmetic complexity, is the content of Section 5. 
 
Next, in Section 6, we consider the ring $\F$ of functions on $\Z^n$. This ring is very important in several branches of engineering such as signal processing, graphics, 3-D printing and digital control. It is isomorphic to the vector space dual $A'$ of $A$, and the automorphism $\alpha$ acts on $\F$ by duality. Thus $(\F,~\alpha)$ is also a difference ring. The results of the ISP for $(A,~\alpha)$ then extend to results for $(\F,~\alpha)$.

When the degree of the automorphism $\alpha$ is bigger than 1, the subspace of summable elements in both $(A,~\alpha)$ and $(\F,~\alpha)$ is infinite dimensional. We topolgise the algebras $A$ and $\F$ in a natural way (as inductive and projective limits, respectively, of finite dimensional Hermitian 
vector spaces) and show that the subspace of summable elements is, however, nowhere dense.

In the last section, we return to the classical ISP of the beginning of this introduction. Now, the group $[\alpha]$ generated by the shift operator on the ring $\F$ of functions on $\Z$ is infinite cyclic, and the results of the previous sections do not apply (of course, it is elementary to solve the ISP for $\F$ directly and to see that every function in it is summable). However, it turns out that when we consider the subring $\F_c$ of functions with finite support, then the solution to the ISP is analogous to the solution when the automorphism is of finite order. As observed above, these solutions are equivalent to the calculation of the low dimensional group cohomologies of $[\alpha]$ with values in $\F$ and in $\F_c$, which are now seen to exhibit a Poincare type of duality.

 \section{automorphisms of the Laurent ring of finte order}
 
 Let $A = \mathbb{C}[\sigma_1, \sigma_1^{-1}, \ldots,\sigma_n, \sigma_n^{-1}]$. It is the $\C$-algebra of shift operators on the lattice $\Z^n$, generated by the forward shifts $\sigma_i: \Z^n \rightarrow\Z^n$, mapping $(x_1,\ldots, x_i, \ldots, x_n)$ to $(x_1,\ldots, x_i+1, \ldots, x_n)$, $i=1,\ldots, n$, together with their inverses, the backward shifts. Let $\F$ be the space of complex valued functions on $\Z^n$. The operator $\sigma_i$ acts on it by $\sigma_i(f)(x) = f(\sigma_i(x))$, for  $f \in \F$ and $x \in \Z^n$.
 
 The set of monomials $B = \{ \sigma_1^{x_1}\cdots \sigma_n^{x_n},~x_i \in \Z, ~1 \leqslant i \leqslant n\}$ is a $\C$-basis for $A$. The number of terms in an element of $ A$ is the number of summands that occur in its expression as a $\C$ -linear sum of the above basis $B$. 
 In what follows, the monomial $\sigma^x = \sigma_1^{x_1}\cdots \sigma_n^{x_n}$
 will be identified with the point $x = (x_1,\ldots, x_n) \in \Z^n$; thus $A$ is isomorphic to the $\C$-vector space spanned independently by the points of $\Z^n$.
  The space $\F$  is then isomorphic to the vector space dual $A'$ of $A$.    
 \vspace{1.5mm}
 
 The group $\aut_\C(A)$ of $\C$-algebra automorphisms of $A$ is described in \cite{PS, sh}. It is the semidirect product $(\C^*)^n \rtimes_\phi \gl$ defined by the homomorphism $\phi: \gl \rightarrow  \aut((\C^*)^n)$, where $\phi(M)(R) = (\prod_{i =1}^n r_i^{m_{1i}}, \ldots, \prod_{i=1}^n r_i^{m_{ni}})$, for $M = (m_{ij})\in \gl$ and $R = (r_1,\ldots, r_n) \in (\C^*)^n$. The automorphism $\alpha=(R,M)$ of $A$, corresponding to $R$ and $M$, is defined by $\alpha(\sigma_i) =  r_i\sigma_1^{m_{i1}} \cdots \sigma_n^{m_{in}}$, $i = 1, \ldots, n$. It maps monomials to monomials (a ring automorphism maps units of the ring to units), and so the number of terms in $\alpha(a)$ is equal to the number of terms in $a$.  
 
 Both $(\C^*)^n$ and $\gl$ will be considered to be subgroups of $\aut_\C(A)$, the first via the canonical inclusion and the second via the splitting $\gl \hookrightarrow (\C^*)^n \rtimes_\phi \gl$ of the exact sequence
$1 \rightarrow (\C^*)^n \longrightarrow (\C^*)^n \rtimes_\phi \gl \longrightarrow \gl \rightarrow 1.$  
 
 \vspace{2mm}
 
Now let $\alpha$ be torsion in $\aut_\C(A)$, i.e. an automorphism of $A$ of finite order, say $d > 1$. The subject of this article is the ISP for the difference ring  $(A,\alpha)$.  

\begin{definition}
The trace with respect to $\alpha$ is the $\C$-linear map $\trace: A \rightarrow A$ defined by $\trace(a) = \sum_{i=0}^{d-1} \alpha^i(a)$. 
\end{definition}

Let $[\alpha]$ be the cyclic group generated by $\alpha$. To determine $\trace$, it is necessary to calculate the trace of all the monomials in the basis $B$, for which we describe their orbits under the action of $[\alpha]$ (\cite{sh} povides a general description of the orbits of monomials under the action of an arbitrary finite subgroup of $\aut_\C(A)$).


\begin{proposition} 
Let the $[\alpha]$-orbit of $\sigma^x \in B$, corresponding to the point $x \in \Z^n$, be  $O<x> ~= ~<\sigma^x, \alpha(\sigma^x), \ldots, \alpha^{d-1}(\sigma^x)>$, and let $A<x>$ denote its $\C$-span in $A$. 
\vspace{1.5mm}

\noindent (i) If the points in $O<x>$ are linearly independent, i.e. if $\dim(A<x>) = d$, then $\trace(\sigma^x) \neq 0$.
\vspace{1mm}

\noindent (ii) If the points in $O<x>$ are linearly dependent, then there are $r \geqslant 2$ points in it that are $\C$-multiples of $\sigma^x$. Let them be  denoted by $O(x) = ~<\sigma^x, \alpha^{i_1}(\sigma^x), \ldots, \alpha^{i_{r-1}}(\sigma^x)>$.

It follows that 

(a) $\trace(\sigma^x) \neq 0$ if and only if every point in $O(x)$ is $\sigma^x$.

(b) $\trace(\sigma^x) = 0$ if and only if for some $j$, $\alpha^{i_j}(\sigma^x) = \zeta \sigma^x$, where $\zeta$ is a root of unity not equal to 1. 

Now, $\dim(A<x>) = \frac{d}{r}.$
\end{proposition}
\noindent Proof: 
\noindent (i) It has already been observed that an algebra automorphism of $A$ maps monomials to monomials. Hence, the points in the orbit $O<x>$ are linearly independent if and only if no two are $\C$-multiples of each other.  Then clearly $\trace(\sigma^x) \neq 0$. 
\vspace{1.5mm}

\noindent (ii) Linear dependence of the points in $O<x>$ implies that there are two points in the orbit that are $\C$-multiples of each other, say $\alpha^{i_m}(\sigma^x) = c\alpha^{i_{m'}}(\sigma^x)$, $m \neq m'$. Then  $\alpha^{i_m - i_{m'}}(\sigma^x) = c\sigma^x$. Let all the points in the orbit that are $\C$-multiples of $\sigma^x$ be denoted $O(x) = ~<\sigma^x, \alpha^{i_1}(\sigma^x), \ldots, \alpha^{i_{r-1}}(\sigma^x)>$, as in the statement.

Denote by $[\alpha](x)$ the subset  $\{1, \alpha^{i_1}, \ldots, \alpha^{i_{r-1}}\}$ of $[\alpha]$ that map $\sigma^x$ to $\C$-multiples of $\sigma^x$. It is a subgroup of $[\alpha]$, and $O(x)$ is the $[\alpha](x)$-orbit of $\sigma^x$. 
Let $\ell = \frac{d}{r}$. It follows that $\{\sigma^x, \alpha(\sigma^x), \ldots, \alpha^{\ell-1}(\sigma^x)\}$ is a maximal set of linearly independent points in the orbit $O<x>$, and that $\alpha^\ell(\sigma^x) = \zeta \sigma^x$, where $\zeta$ is an $r$-th root of unity.

Let $[\alpha]_x$ be the stabilizer of $x$, and let it contain $s$ elements. The action of the quotient $[\alpha](x)/[\alpha]_x$ on $\sigma^x$ is then isomorphic to the action given by multiplying $\sigma^x$ by the $k$-th roots of unity, where $k = \frac{r}{s}$ (thus, the $\zeta$ in the above paragraph is a primitive $k$-th root of unity).  Each point in $O(x)$ occurs the same number of $s$ times in $O<x>$.  

Let $c\sigma^{x'} = \alpha^i(\sigma^x)$ be any point in $O<x>$ not in $O(x)$. Then the subgroup $[\alpha](x')$ of elements of $[\alpha]$ that map $\sigma^{x'}$ to multiples of itself, is equal to $[\alpha](x)$, its stabilizer $[\alpha]_{x'}$ is equal to $[\alpha]_x$, and hence the action of $[\alpha](x')/[\alpha]_{x'}$ on $c\sigma^{x'}$ is again
isomorphic to the action given by multiplying  $c\sigma^{x'}$ by the same $k$-th roots of unity, as above. It now follows that every element in the orbit $O<x>$ of $\sigma^x$ occurs the same number of times as any other element in it.
\vspace{1.5mm}

Suppose first that each point in $O(x)$ is $\sigma^x$, i.e. suppose that that the stabilizer $[\alpha]_x$ equals $[\alpha](x)$. 
Then $r\sigma^x$ is a summand of $\trace(\sigma^x)$, and as the other summands do not involve $\sigma^x$, it follows that $\trace(\sigma^x) \neq 0$.

\vspace{1.5mm}
On the other hand, if $[\alpha]_x \subsetneq [\alpha](x)$, 
then one summand of $\trace(\sigma^x)$ equals $s(1 + \zeta + \cdots + \zeta^{k-1})\sigma^x$, where $\zeta $ is a primitive $k$-th root of unity, $k > 1$, and is hence equal to 0. The other summands in $\trace(\sigma^x)$ are similarly 0. Hence $\trace(\sigma^x) = 0$.
\hspace*{\fill}$\square$\\ 

Let $\{A_i ~|~ i\in I\}$ be the set of all the subspaces $A<x>$ determined by the orbits of all the monomials $\sigma^x$ in $B$, as in Proposition 2.1. Then, $A = \bigoplus_{i\in I} A_i$. 
Each $A_i$ is $[\alpha]$-invariant, and is a finite dimensional representation of $[\alpha]$. 

Suppose that the summand $A_i$ is the subspace $A<x>$ equal to the $\C$-span of the orbit $O<x>$ of $\sigma^x$. It decomposes into a direct sum $A_{i'} \oplus A_{i''}$ of representations of $[\alpha]$, where $A_{i'}$ is the trivial representation on the subspace spanned by $\trace(\sigma^x)$, and  $A_{i''}$ is the standard represention on the subspace $\{\sum_{i=0}^{d-1} c_i\alpha^i(\sigma^x)~|~ c_i \in \C,  ~\sum_i c_i = 0\}$, (for instance \cite{ser}). 
\vspace{2mm}

Define $A_{'} = \bigoplus_i A_{i'}$ and $A_{''} = \bigoplus_i A_{i''}$; thus $A = A_{'} \oplus A_{''}$.

\begin{lemma} 
The elements of $A_{''}$ have zero trace, whereas the nonzero elements of $A_{'}$ have nonzero trace.
\end{lemma}
\noindent Proof: 
If $a \in A_{i''}$, then $a = \sum_{i=0}^{d-1} c_i\alpha^i(\sigma^x)$, where $\sum_i c_i = 0$. It follows that $\trace(a) = \sum_{j=0}^{d-1} \alpha^j(a) = \sum_{i,j=0}^{d-1} c_i \alpha^{i+j}(\sigma^x) = \sum_i c_i \trace(\alpha^i(\sigma^x)) = \trace(\sigma^x) \sum_i c_i = 0$, as $\trace(\alpha^i(\sigma^x)) = \trace(\sigma^x)$.

Let $a \in A_{i'}$; then it is a scalar multiple of $\trace(\sigma^x)$ for some $\sigma^x \in B$. As $\trace \circ \trace = d\trace$, 
$\trace(a) \neq 0$ if $\trace(\sigma^x) \neq 0$.
\hspace*{\fill}$\square$

\section{solution to the ISP for the Laurent ring}

Let $\alpha$ be a $\C$-algebra automorphism of $A$ of order $d > 1$, and consider the linear map $(\alpha - 1): A \rightarrow A$. Its image is the set of summable elements in $(A, \alpha)$; it is an $\alpha$-invariant $\C$-vector subspace of $A$ that is closed under multiplication by constants.

As $(\alpha^i - 1) = (\alpha - 1)(1 + \alpha + \cdots +\alpha^{i-1})$, the image of $(\alpha - 1)$ contains the image of $(\alpha^i - 1):A \rightarrow A$, for $1 \leqslant i \leqslant d-1$; hence the space of $[\alpha]$-coinvariants of $A$, $A_{[\alpha]} = A/<\alpha^ia - a,~ 1 \leqslant i \leqslant d-1, a \in A> $, is equal to the cokernel $A/\im(\alpha - 1)$ of $(\alpha - 1)$.
\vspace{1mm}

The kernel of $(\alpha-1)$, viz. the ISP  when $b = 0$, is the constant subring of $(A, \alpha)$. Given one solution to $(\alpha-1)a = b$, every other solution is determined by adding a constant to it. 
As $[\alpha]$ is cyclic, the set of constants of $(A,~\alpha)$ is also the set $A^{[\alpha]}$ of $[\alpha]$-invariants of $A$. It is the zero-th group cohomology $H^0([\alpha], A)$, and is described in Section 4.

\vspace{1.5mm}
The composition $\trace \circ ~(\alpha - 1)$ is equal to 0, hence the image of $\alpha - 1$ is contained in the kernel of $\trace$. The following result (as in Hilbert Theorem 90 in \cite{lang}) shows that this containment is an equality. 

\begin{theorem} 
An element $b$ in $A$ is summable in $(A,\alpha)$ if and only if 
$\trace(b) = 0$. Thus, the space of summable elements is infinite dimensional.
\end{theorem}
\noindent Proof: Suppose that $\trace(b) = 0$. 
Then $a = \frac{-1}{d}(b + (b + \alpha(b)) + \cdots + (b + \alpha(b) + \cdots + \alpha^{d-2}(b)))$ 
satisfies $\alpha(a) - a = b$.

In the direct sum decomposition $A = A_{'} \oplus A_{''}$ of Lemma 2.1, the elements of $A_{''}$ have zero trace, and the nonzero elements of $A_{'}$ have nonzero trace. Thus, the space of summable elements is precisely the subspace $A_{''}$. It now suffices, in the notation preceeding Lemma 2.1, to show that there are infinitely many indices $i$ in $I$ such that $A_{i''}$ is nonzero. 

Suppose that $A_i$ is the space $A<x>$ determined by the orbit $O<x>$ of $\sigma^x \in B$. In the notation of Proposition 2.1, when the points in $O<x>$ are linearly independent, $\dim(A_i) = d$, and  $\dim(A_{i''})$, the dimension of the standard representation, equals $d-1$. When the points are dependent and $\trace(\sigma^x) \neq 0$, then $\dim(A_i) = \frac{d}{r} = \ell$, and $\dim(A_{i''}) = \ell - 1$, where $r$ is the number of points in $O<x>$ that are multiples of $\sigma^x$ (here, these $r$ points are all $\sigma^x$ as trace is nonzero). Finally, when $\trace(\sigma^x) = 0$, $\dim(A_i) = \dim(A_{i''}) = \ell$.

Thus, an $A_i$ does not contribute to the space of summable elements if and only if the points in $O<x>$ are linearly dependent, $\trace(\sigma^x) \neq 0$ and $r = d$. This is to say that $\alpha^i(\sigma^x) = \sigma^x$ for all $i$, i.e. $\sigma^x$ is a constant. The set of all constants in $B \subset A$, corresponding to points in $\Z^n$, is a sublattice $\mathbb{S}$ of $\Z^n$. If $\mathbb{S} = \Z^n$, then $\alpha$ is the identity, contadicting the assumption that the degree of $\alpha$ is equal to $d > 1$. Thus there are infinitely many points in $x \in \Z^n \setminus \mathbb{S}$, and the space of summable elements is infinite dimensional.  
\hspace*{\fill}$\square$\\

\noindent Remark 3.1: We can write down an explicit basis for the space of summable elements by writing down one for each $A_{i''},~ i \in I$.

Suppose that an $A_i$ is the space determined by the orbit $O<x>$ of $\sigma^x$,  and is of dimension $\frac{d}{r} = \ell$. As observed in the proof of Proposition 2.1, $\{\sigma^x, \alpha(\sigma^x), \ldots, \alpha^{\ell-1}(\sigma^x)\}$ is a basis for $A_i$, and  $\alpha^\ell(\sigma^x) = \zeta \sigma^x$, $\zeta$ a root of unity.

\vspace{1.5mm}
Suppose $\trace(\sigma^x) \neq 0$. If $\ell = 1$, then $\sigma^x$ is a constant and $A_{i''} = 0$.

Suppose that $\ell > 1$. As trace is nonzero, it follows that 
$\alpha^\ell(\sigma^x) = \sigma^x$.

Let $\mu$ be an $\ell$-th root of unity, not equal to 1. Then $\{\sum_{i=0}^{\ell - 1} \mu^{-ij}\alpha^i(\sigma^x), ~ 1\leqslant j \leqslant \ell - 1\}$, is a basis for the space of summable elements in $A_i$.

\vspace{1mm}
If $\trace(\sigma^x) = 0$, then every point in $O<x>$ is summable, and $\{\sigma^x, \alpha(\sigma^x), \ldots, \alpha^{\ell-1}(\sigma^x)\}$  is a basis for the space of summable elements in $A_i$. 
\\

\noindent Remark 3.2: The composition $(\alpha - 1) \circ \trace$ is also equal to 0. Let $a \in \ker(\alpha - 1) = A^{[\alpha]}$. Then $\trace(\frac{a}{d}) = a$, and thus the the image of $\trace$ is equal to $A^{[\alpha]}$.  
This image is disjoint from the subspace of summable elements, because $\trace \circ \trace = d\trace$, as we have observed above in the direct sum decomposition $A = A_{'} \oplus A_{''}$. 

We also observe directly a general fact about coinvariants and invariants of finite group actions:
$A_{[\alpha]} = \coker(\alpha - 1) \simeq A/\ker(\trace) \simeq \im(\trace) =  A^{[\alpha]}$. 
\hspace*{\fill}$\square$\\  

We specialise to the case when the automorphism $\alpha$ belongs to either factor of $\aut_\C(A)$. In the notation described in the beginning of Section 2, an element $(R, M) \in \aut_\C(A)$ is in $\gl$ if $R = (1, \cdots, 1)$ in $(\C^*)^n$, and is in $(\C^*)^n$ if $M$ is the identity matrix in $\gl$. 

\begin{corollary} Let $\alpha$ be an automorphism of $A$ of order $d > 1$.

\vspace{1mm}
\noindent (i) Suppose $\alpha \in \gl$. Then $\trace(\sigma^x) \neq 0$ for every $\sigma^x \in B$, and hence, for every $i \in I$ (in the notation preceeding Lemma 2.1), the space of summable elements in $A_i$ is a codimension 1 subspace of it.

\vspace{1mm}

\noindent (ii) Suppose $\alpha = (r_1, \ldots, r_n) \in (\C^*)^n$. Then the monomial $\sigma^x = \sigma_1^{x_1} \cdots \sigma_n^{x_n}$ (corresponding to $x = (x_1, \ldots, x_n) \in \Z^n$) satisfies $\trace(\sigma^x) \neq 0$ if and only if $r_1^{x_1} \cdots r_n^{x_n} = 1$. Such monomials are constants of $(A, \alpha)$ and correspond to points in a proper sublattice of $\Z^n$. The corresponding indices $i \in I$ satisfy $\dim(A_{i{''}}) = 0$. For indices $i$ corresponding to points in $\Z^n$ outside this sublattice, $\dim(A_{i{''}}) = 1$ 

An $a \in A$ is summable if and only if each of its monomial terms is summable.
\end{corollary}
\noindent Proof: (i) The automorphism $\alpha$ now maps $B$ to itself. Hence the trace of every $\sigma^x$ in $B$ is nonzero, viz. the calculation in the proof of Theorem 3.1. This calculation also establishes the claim on the dimension of $A_{i''}$.

\vspace{1mm}
\noindent (ii) The action of $\alpha$ on $\sigma^x = \sigma_1^{x_1} \cdots \sigma_n^{x_n}$ is to multiply it by $r^x = r_1^{x_1} \cdots r_n^{x_n}$; hence it follows that $\alpha^i(\sigma^x) = (r^x)^i \sigma^x$. The points $x \in \Z^n$, corresponding to $\sigma^x$ such that $r^x = 1$, is a proper sublattice of $\Z^n$, as the $\deg(\alpha) = d > 1$.

The other assertions all follow from the calculation in Theorem 3.1.
\hspace*{\fill}$\square$\\

\noindent Remark 3.3: By a theorem of Minkowski \cite{mink}, the order of any finite subgroup of $\gl$ is bounded by a constant (depending only on $n$), and which, therefore, also bounds the orders of all its torsion elements. Let the set of these possible orders be $D = \{1, \ldots, d_r \}$.

\vspace{1mm}
Let $\alpha = (\alpha_1, \alpha_2)$ be a torsion element in $\aut_\C(A) = (\C^*)^n \rtimes_\phi \gl$ of degree $d$. Let the order of $\alpha_2 \in \gl$  be $d_i$ in $D$. It follows that $d$ is divisible by $d_i$. 

Thus, if $d$ is not divisible by any element in $D$ other than 1, it follows that $\alpha_2 = 1$, and that $\alpha$ is in $(\C^*)^n$. Then, the space of solutions to the ISP is described by (ii) of the above corollary. \hspace*{\fill}$\square$\\

The set of summable elements in $(A, \alpha)$ is an infinite dimensional proper subspace of $A$. We equip $A$ with the structure of a topological vector space to make a more precise statement as follows. 

Given $\alpha \in \aut_\C(A)$ of order $d > 1$, we topologise $A$ as follows: 
in the notation at the end of Section 2, $A = \bigoplus_{i\in I} A_i$ is the direct sum decomposition of $A$ into $[\alpha]$-invariant subspaces determined by orbits of monomials in $B$.
For every $i$, let $\mathcal{A}_i$ denote the topological vector space given by a Hermitian inner product on $A_i$. The automorphism $\alpha$ restricts to a continuous map on $\mathcal{A}_i$. 
Let $\mathfrak{D}$ be the directed system whose objects are finite sums $\{\mathcal{A}_{i_1} \oplus \cdots \oplus \mathcal{A}_{i_t} ~|~ i_1, \ldots, i_t \in I\}$, and whose morphisms are inclusions.
We denote by $\mathcal{A}$ the inductive limit topology on $A$ given by $\mathfrak{D}$. The space $\mathcal{A}$ is then a separable (strict) $LB$-space, i.e. the inductive limit of a countable inductive system of Banach spaces. 
\begin{proposition} Let $\alpha \in \aut_\C(A)$ be of finite order. Then the infinite dimensional subspace of summable elements in $(A, \alpha)$ is proper closed in $\mathcal{A}$, and hence nowhere dense in it.
\end{proposition}
\noindent Proof: The map $\alpha: \mathcal{A} \rightarrow \mathcal{A}$
is a bounded linear map, hence continuous. 
Thus, $\trace: \mathcal{A} \rightarrow \mathcal{A}$ is also continuous. The space of summable elements is its kernel, and is thus closed in $\mathcal{A}$.
\hspace*{\fill}$\square$\\

\noindent Remark 3.3: Any other $\alpha' \in \aut_\C(A)$ of finite order defines a directed system which is cofinal with respect to that for $\alpha$, and $\mathcal{A}$ is thus independent of $\alpha$.\\

\section{$H^i([\alpha], A), ~i = 0, 1$}

Consider the difference ring $(A, \alpha)$ as an $[\alpha]$-module, where as before, $\alpha$ is an automorphism of $A$ of order $d$. The computations in the solution of the ISP also calculate, {\it ab initio}, the group cohomologies $H^0([\alpha], A)$ and $H^1([\alpha], A)$. 

The cohomology groups $\{H^k([\alpha], A), k \geq 0\}$ of $[\alpha]$ with coefficients in $A$ is the cohomology of the complex $0 \rightarrow C^0 \stackrel{\delta}{\rightarrow} \cdots C^{k-1} \stackrel{\delta}{\rightarrow} C^k \stackrel{\delta}{\rightarrow} C^{k+1} ~\cdots$, where $C^k$ is the group $\{\theta: [\alpha]^k \rightarrow A\}$ of $k$-cochains, and $\delta(\theta)(\alpha^{i_1}, \ldots, \alpha^{i_{k+1}}) =     \alpha^{i_1}\theta(\alpha^{i_2}, \ldots, \alpha^{i_{k+1}}) + \sum_{j=1}^k (-1)^j\theta(\alpha^{i_1}, \ldots, \alpha^{{i_j}+i_{j+1}}, \ldots, \alpha^{i_{k+1}}) + (-1)^{k+1}\theta(\alpha^{i_1}, \ldots, \alpha^{i_k})$.      

In particular, $C^0 = \{\theta: [\alpha]^0 \rightarrow A\}$ can be identified with points of $A$. An $a \in A$ is in $H^0([\alpha], A)$ if $\delta(a): [\alpha] \rightarrow A$ equals 0. This is to say that $\delta(a)(\alpha^i) = \alpha^i(a)-a = 0$ for all $i$, and thus that $a$ is in $A^{[\alpha]}$, the set of $[\alpha]$-invariants of $A$.

\begin{proposition} $H^0([\alpha], A)$ is equal to the constant subring of $(A, ~\alpha)$, and is infinite dimensional.
\end{proposition}
\noindent Proof: The set of constants is the set of $[\alpha]$-invariants of $A$. Thus, $H^0([\alpha], A)$ equals the constant subring of $A$.

\vspace{1mm}

In the notation of Proposition 2.1, the dimension of $H^0$ is equal to the dimension of $A_{'} = \bigoplus_i A_{i'}$, the space of the trivial representation. By the calculation in Theorem 3.1, if $A_i$ is $A<x>$, the space spanned by the orbit $O<x>$, then $\dim{A_{i'}} = 1$ when the points in $O<x>$ are linearly independent. When the points are dependent and $\trace(\sigma^x) \neq 0$, then again $\dim(A_i') = 1$, but when $\trace(\sigma^x) = 0$, then $\dim(A_i') = 0$.   

There is nothing to be done if the trace of every $\sigma^x$ is nonzero. Hence suppose that there is an $\sigma^x \in B$ whose trace is equal to 0. 
It then follows, by the proof of Proposition 2.1, that the set of those points in  $O<x>$ which are dependent on $\sigma^x$ is $\{\sigma^x, \zeta \sigma^x, \ldots, \zeta^{k-1} \sigma^x\}$, where $\zeta$ is a primitive $k$-th root of unity, for some $k > 1$. Consider the points $\{jkx \in \Z^n ~|~  j \geqslant 1\}$ corresponding to the monomials $\sigma^{jkx} \in B$. The trace of each of them is nonzero, and hence there are infinitely many indices $i \in I$ such that $\dim(A_{i'}) > 0$. This proves that $H^0([\alpha], A)$ is infinite dimensional. 
\hspace*{\fill}$\square$\\
 
\begin{corollary} There are infinitely many solutions $a$ to $\alpha(a) - a = b$ when $\trace(b) = 0$. 
\end{corollary}

 \begin{proposition} $H^1([\alpha], A) = 0$.
 \end{proposition}
 \noindent Proof: A 1-cochain $\theta$ is a cocycle if $\delta(\theta)(\alpha^i, \alpha^j) = \alpha^i \theta(\alpha^j) - \theta(\alpha^{i+j}) + \theta(\alpha^i) = 0$, for all $i, j$. This implies that $\theta(1) = 0$, and that $\theta(\alpha^i) = \sum_{j=0}^{i-1} \alpha^j \theta(\alpha)$ for $i \geqslant 1$. Then, $\trace(\theta(\alpha)) =  \sum_{j=0}^{d-1} \alpha^j \theta(\alpha) = \theta(\alpha^d) = 0$. By Proposition 3.1, there is an $a \in A$ such that $\theta(\alpha) = \alpha(a) - a$. It now follows that $\theta(\alpha^i) = \alpha^i(a) - a$, and hence that $\theta = \delta(a)$ is a coboundary.
 \hspace*{\fill}$\square$\\
 
\noindent Remark 4.1: For the finite cyclic group $[\alpha]$, tensoring the standard 2-periodic free resolution of $\Z$ over the group ring $\Z[\alpha]/(\alpha^d - 1)$ by $A$, calculates all the $H^i([\alpha], A)$, (for instance \cite{b}). The point of the above proposition was only to show that the calculations involved in the ISP also calculate $H^1$. 

Furthermore, it follows from this standard resolution that $H^2([\alpha], A) = A^{[\alpha]}/\trace(A)$. By the observation in Remark 3.1, $H^2 = 0$. 
This resolution also implies that the group cohomologies of a finite cyclic group are 2-periodic, and it follows that $H^i([\alpha], A) = 0$ for all $i > 0$. 
 
\section{a finite procedure} 
 
 Let $\alpha \in \aut_\C(A)$ be an element of degree $d$. In the notation at the end of Section 2, let $A = \bigoplus_{i\in I} A_i$ be the direct sum decomposition of $A$ into $[\alpha]$-invariant subspaces determined by orbits of monomials in the basis $B$. let $\alpha_i$ denote $\alpha$ restricted to $A_i$. Given $b \in A$, let $b = \sum b_i$, where $b_i \in A_i$. Then, $b$ is summable in $(A, \alpha)$ if and only if each $b_i$ is summable in $(A_i, \alpha_i)$, and this can be reduced to a linear algebra calculation.
 
 Let $(\alpha_i-1)$ be the matrix of $\alpha_i-1$ with respect to some basis for $A_i$, of dimension $k_i$, say. Similarly, let $(\tracei)$ be the matrix of $\trace$ restricted to $A_i$. Then the sequence
 \begin{equation}
 \C^{k_i} \stackrel{(\alpha_i - 1)}{\longrightarrow} \C^{k_i} \stackrel{(\tracei)} {\longrightarrow} \C^{k_i}
 \end{equation}
 is exact by Proposition 3.1 (the rows of $(\tracei)$ span the subspace of all relations between the $k_i$ rows of $(\alpha_i-1)$). 
 Thus to decide whether $b_i \in A_i$ is summable, it is necessary and sufficient to check whether $(\tracei)b_i =0$. As only finitely many $b_i$ in $b = \sum_ib_i$ are nonzero, this check provides a finite procedure for the solution of the ISP for the Laurent ring.
 
 \vspace{1.5mm}
 The dimension $k_i$ of $A_i$ is bounded by the order $d$ of $\alpha$. This implies that the arithmetic complexity of determining when an element of $A_i$ is summable is {\it a priori} $O(d^2)$ (counting the complexity of multiplication as $O(1)$). Results of Sections 2 and 3 calculate the dimension of $A_i = A<x>$ in terms of the orbit of the corresponding monomial $\sigma^x$, and this allows us to improve this estimate. In particular, Corollary 3.1 establishes the following special cases.
 
 \begin{proposition} Let $\alpha \in \aut_\C(A)$ be of order $d$. Let $r$ be the number of points in the $[\alpha]$-orbit $O<x>$ of $\sigma^x = \sigma_1^{x_1} \cdots \sigma_n^{x_n}$ that are dependent on it (in the notation of Proposition 2.1).
 
 \noindent (i) Let $\alpha \in \gl$. Then the complexity of determining whether $b \in A<x>$ is summable is $O((\frac{d}{r})^2)$. Thus, if the points in $O<x>$ are linearly independent, then this complexity is $O(d^2)$.
 
 \noindent (ii) Let $\alpha = (r_1, \ldots, r_n) \in (\C^*)^n$. Then the complexity of determining whether $b \in A<x>$ is summable is $O(n + \sum_{i=1}^n \log|x_i|)$. 
 \end{proposition}
 \noindent Proof: (i) requires multiplying a square matrix and a column vector  
 of size $\frac{d}{r}$, and checking whether the entries of the product are all 0. If $r = d$, then $x$ is a fixed point of the action, and there are no nonzero summable elements in the 1 dimensional span $A<x>$.
 
 In (ii), it suffices to multiply the powers $r_i^{x_i}, i = 1, \ldots, n$. If this product equals 1, then $x$ is a fixed point, and there are no nonzero summable elements in $A<x>$. Otherwise, every element in it is summable. The complexity of the operation of evaluating $n$ powers and $n-1$ multiplications is $O(n + \sum_{i=1}^n \log|x_i|)$, \cite{knuth}.
\hspace*{\fill}$\square$\\ 
 
 \noindent Example: Let $A =\C[\sigma_1, \sigma_1^{-1}, \sigma_2, \sigma_2^{-1}]$ be the ring of shift operators on $\Z^2$, and let 
 $\alpha$ in $\aut_\C(A)$ be defined by $\alpha(\sigma_1) = -\sigma_2^{-1}$, $\alpha(\sigma_2) = \sigma_1^{-1}$. The degree of $\alpha$ equals 4. 
 
 \vspace{1mm}
 \noindent (i) Let $A_1$ be the $\alpha$-invariant subspace of $A$ spanned by $ \{\sigma_1, \sigma_1^{-1}, \sigma_2, \sigma_2^{-1}\}$.  To determine if an element in $A_1$ is summable in $(A, \alpha)$, it needs to be determined whether it is summable in $(A_1, \alpha|_{A_1})$. The matrix of $(\alpha - 1)|_{A_1}$ with respect to the above basis is
 \[
 (\alpha_1-1)  = \left(
\begin{array}{lccc}
 
-1 & \phantom{-}0 & \phantom{-}0 & \phantom{-}1 \\
\phantom{-}0 & -1 & \phantom{-}1 & \phantom{-}0 \\
\phantom{-}0 & -1 & -1 & \phantom{-}0 \\
-1 & \phantom{-}0 & \phantom{-}0 & -1
 \end{array}
\right).
\]
The corresponding $\traceo$ equals 0, and thus $(\alpha - 1)|_{A_1}$ is surjective. Every element in $A_1$ is summable; for instance $\sigma_1 = (\alpha - 1)(\frac{-\sigma_1}{2} + \frac{\sigma_2^{-1}}{2})$.

\vspace{1mm}
\noindent (ii) Consider the $\alpha$-invariant subspace $A_2$ of $A$ spanned by $ \{\sigma_1\sigma_2, ~\sigma_1^{-1}\sigma_2^{-1}\}$. 
The matrix of $(\alpha - 1)|_{A_2}$ with respect to this basis is
 \[
 (\alpha_2-1)  = \left(
\begin{array}{lc}
 
 -1 & -1 \\
 -1 & -1
 \end{array}
\right),
\]
and the corresponding $(\tracet)$ equals $\left(
\begin{array}{lc}
 
 \phantom{-}1 & -1 \\
 -1 & \phantom{-}1
 \end{array}
\right)$.

\vspace{1.5mm}

As $
(\tracet) \left( \begin{array}{l} 1 \\ 0 \end{array} \right) \neq 0
$
it follows that $\sigma_1\sigma_2$ is not summable.

On the other hand, $\sigma_1\sigma_2 + \sigma_1^{-1}\sigma_2^{-1}$ is in the kernel of $\tracet$, and is hence summable. It equals $(\alpha - 1)(-\frac{1}{2}(\sigma_1\sigma_2 + \sigma_1^{-1}\sigma_2^{-1}))$.  
\hspace*{\fill}$\square$\\

\section{ISP for the ring of functions $\F$}
Recall from Section 2 the ring $\F = \{f: \Z^n \rightarrow \C\}$ of complex valued functions on the lattice $\Z^n$. As before, we identify the monomial $\sigma^x$ in the basis $B$ with point $x$ in $\Z^n$. Thus, an $f \in \F$ extends linearly to a unique element in the dual $A'$, also denoted $f$. Conversely, an element in $A'$, upon restriction to $\Z^n$, corresponds to a unique element in $\F$.  

An element $\alpha$ of $\aut_\C(A)$ defines an automorphism of $A'$ by $\alpha(f)(a) = f(\alpha^{-1}(a))$, for $f \in A'$ and $a \in A$. By the above correspondence, i.e. by restriction to points in $\Z^n$, $\alpha$ is an automorphism of $\F$. 
The pair $(\F, \alpha)$ is a difference ring, and the Indefinite Summation Problem now is: given $g \in \F$, decide whether there exists an $f \in \F$ such that $\alpha(f) - f = g$. 

Let $\alpha$ be of finite order $d$.  Define $\tracep: A' \rightarrow A'$ by $\tracep(f) = \sum_{i=0}^{d-1} \alpha^i(f)$. 
Again, restricting to points in $\Z^n$, we have $\tracep: \F \rightarrow \F$.

\vspace{1.5mm}
We  use the results in Section 3 to solve the IPS for $(\F, \alpha)$.

\begin{theorem} 
Let $\alpha \in \aut_\C(A)$, of order $d$, act on $\F$ by duality. Then $g \in \F$ is summable in $(\F, \alpha)$ if and only if $g$ is in the annihilator $(A_{'})^o$ of the subspace $A_{'}$ of $A$ of elements of nonzero trace. The space of summable functions is infinite dimensional.
\end{theorem} 
\noindent Proof: Just as in Proposition 3.1, it follows that a $g \in \F$ is summable if and only if $\tracep(g) = 0$, which is to say that $g(\sum_{i=0}^{d-1} \alpha^i(\sigma^x)) = g(\trace(\sigma^x)) = 0$, for every $x \in \Z^n$. 

Let $\sigma^x = \sigma^{x'} + \sigma^{x''}$, where $\sigma^{x'} \in A_{'}$ is of nonzero trace, and $\sigma^{x''} \in A_{''}$ is of trace 0. Then, $g(\trace(\sigma^x)) = g(\trace(\sigma^{x'}))$; hence to be summable, it is necessary and sufficient that $g(\trace(\sigma^{x'})) = 0$ for every $\sigma^{x'} \in A{'}$.

For every index $i$, $A_{i'}$ is the span of some $\trace(\sigma^{x'})$. It now follows that $g \in \F$ is summable if and only if $g$ is in the annihilator $(A_{'})^o$ of $A_{'}$. 

The space of summable functions is therefore infinite dimensional because $A_{''}$ is by Theorem 3.1 
\hspace*{\fill}$\square$\\

Analogous to Corollary 3.1, we have the following.
\begin{corollary} (i) If $\alpha \in \gl$, then $g \in \F$ is summable if and only it vanishes at $\trace(\sigma^x)$, for every $x \in \Z^n$. 
\vspace{1mm}

(ii) If $\alpha = (r_1, \ldots, r_n) \in (\C^*)^n$, then $g$ is summable if and only if it vansishes on the sublattice of points $(x_1, \ldots,x_n) \in \Z^n$ such that $r_1^{x_1} \cdots r_n^{x_n}$ = 1. 
\end{corollary}

\noindent Remark 6.1: Any two solutions to a summable $g$ differ by a constant of $(\F, \alpha)$, and we briefly describe them. A constant $f$
satisfies $f(\alpha^i(x)) = f(x)$, for all $x$ in $\Z^n$ and all $i$. 
This implies the following: if $\trace(\sigma^x) = 0$, then by Proposition 2.1, the orbit through $\sigma^x$ contains $\zeta \sigma^x$, where $\zeta$ is a root of unity not equal to 1. It follows that $f(x)$ must equal $\zeta f(x)$, and hence that $f(x) = 0$. Thus, a constant in $(\F, \alpha)$ must vanish at every $x \in \Z^n$ whose trace equals 0, and hence is in $(A_{''})^o$. 

In other words, if  $f \in (A')^{[\alpha]}$, then $\alpha^i(f) = f$ for all $i$, and $f$ vanishes on $(\alpha - 1)(A)$. Thus $f \in (A_{[\alpha]})' \simeq (A^{[\alpha]})'$ (viz. Remark 3.2), and $H^0([\alpha], ~\F) \simeq H^0([\alpha], ~A)'$. It follows that if $\tracep(g) = 0$, then there are infinitely many solutions $f$ to $\alpha(f) - f = g$ (viz. Proposition 4.1).
\hspace*{\fill}$\square$\\

To make a statement analogous to Proposition 3.2, we consider the inverse system $D'$ dual to the direct system $D$ of Section 3. Its objects are $\{\mathcal{A}'_{i_1} \times \cdots \times \mathcal{A}'_{i_t} ~|~ i_1, \ldots, i_t \in I\}$, and morphisms are restrictions (i.e. dual to the inclusions in $D$). We denote by $\mathcal{A}'$ the projective limit topology on $A'$ given by $D'$ (as $\mathcal{A}$ is the inductive limit of finite dimensional spaces, its continuous dual coincides with its algebric dual).  Being the dual of an $LB$-space, $\mathcal{A}'$ is Fr\`echet \cite{gr}. 

\begin{proposition} Let $\alpha \in \aut_\C(A)$ be of finite order. Then the infinite dimensional subspace of summable elements in $(A', \alpha)$ is proper closed in $\mathcal{A}'$, and hence nowhere dense in it.
\end{proposition}
\noindent Proof: Just as in Proposition 3.2, 
$\tracep: \mathcal{A}' \rightarrow \mathcal{A}'$ 
is continuous. 
The space of summable elements is its kernel, and is thus closed in $\mathcal{A}'$.
\hspace*{\fill}$\square$\\

\section{the classical ISP, revisited}
We return briefly to the classical ISP, namely $(\F, ~\alpha)$ of the Introduction, where $\F$ is the ring $\{f: \Z \rightarrow \C\}$ of functions on $\Z$, and the ring automorphism $\alpha: \F \rightarrow \F$ is the translate $\alpha(f)(x) = f(x+1)$. The group generated by $\alpha$ is the infinite group $[\alpha] = \{\alpha^i ~|~ i \in \Z\}$, and none of the results of the previous sections are applicable now. Indeed, Definition 2.1 of trace carried over here, namely $\trace(f) = \sum_{i \in \Z} \alpha^i(f)$, for $f \in \F$, itself does not make sense.  

\vspace{1.5mm}
Suppose $\F_c \subset \F$ is the subset of functions with {\it finite} support. This subset does not have a multiplicative identity, but otherwise satisfies the other axioms of a commutative ring. The automorphism $\alpha$ maps $\F_c$ to itself, and we can consider the difference `ring' $(\F_c, ~\alpha)$. 

Now $\trace|_{\F_c}: \F_c \rightarrow \F$ is well defined, as $\trace(f)(x) = \sum_{i \in \Z} \alpha^i(f)(x)$ reduces to a finite sum. The image of $\trace|_{\F_c}$ is the set of constant functions on $\Z$, and thus $\trace(f)$ is in $\F_c$ if and only if it equals 0.

Perhaps curiously, vanishing trace is also the condition for summability within $\F_c$, just as in Theorem 3.1

\begin{proposition} An element $g \in \F_c$ is summable in $(\F_c, ~\alpha)$ if and only if $\trace(g) = 0$. Thus, the space of summable functions is infinite dimensional.
\end{proposition}
\noindent Proof: If for some $f \in \F_c$, $g = \alpha(f) - f$, then $\trace(g) = 0$.

Conversely, if $\trace(g) = 0$, then $f: \Z \rightarrow \C$, defined by $f(x) = \sum_{i = -\infty}^{x-1} g(i)$, is in $\F_c$, and satisfies $g(x) = f(x+1) - f(x)$. \hspace*{\fill}$\square$\\

Just as in Section 4, the above solution to the ISP also calculates the low dimensional cohomologies of the group $[\alpha]$ with values in $\F_c$.  

\begin{proposition}  $H^0([\alpha], ~\F_c) = 0$, 

\hspace{2.7cm} $H^1([\alpha], ~\F_c) \simeq \C$.
\end{proposition}
\noindent Proof: There are no nonzero constant functions in $(\F_c, ~\alpha)$, hence $H^0([\alpha], ~\F_c) = 0$. (Thus, the solution $f$ to $\alpha(f) - f = g$, when $\trace(g) = 0$, is unique.) 
\vspace{1mm}

A 1-cochain $\theta: [\alpha] \rightarrow \F_c$ is a cocycle if $\delta(\theta)(\alpha^i, \alpha^j) = \alpha^i\theta(\alpha^j) - \theta(\alpha^{i+j}) + \theta(\alpha^i) = 0$, for all $i,j \in \Z$. This implies that (i) $\theta(\alpha^0) = 0$, and that (ii) $\theta(\alpha^{-i}) = - \alpha^{-i}\theta(\alpha^i)$. It also follows, just as in Proposition 4.2, that (iii) $\theta(\alpha^i) = \sum_{j=0}^{i-1} \alpha^j\theta(\alpha)$, for all $ i \geqslant 1$. Thus, if $\theta$ is a cocycle, then $\theta(\alpha)$ determines the values $\theta(\alpha^i)$ for all $i \in \Z$.

If the cocycle $\theta$ is also a coboundary, say $\theta = \delta(f)$ for some 0 cochain $f \in \F_c$, then $\theta(\alpha^i) = \alpha^i(f) - f$, for all $i \in \Z$. In particular, $\theta(\alpha)$ is summable in $\F_c$, and hence $\trace(\theta(\alpha)) = 0$ by the above proposition.

Conversely, suppose $\trace(\theta(\alpha)) = 0$, so that $\theta(\alpha) = \alpha(f) - f$ for some $f \in \F_c$ (again by the above proposition). Then it follows from (i) and (iii) above that $\theta(\alpha^i) = \alpha^i(f) - f$ for all $i \geqslant 0$, and then by (ii) that $\theta(\alpha^i) = \alpha^i(f) - f$ also for negative $i$. Thus, $\theta = \delta(f)$ is a coboundary if and only if $\trace(\theta(\alpha)) = 0$. As trace can be any complex number, the proposition follows.
\hspace*{\fill}$\square$\\

The ISP can be directly solved for $(\F, ~\alpha)$.

\begin{proposition} Every $g \in \F$ is summable in $(\F, ~\alpha)$, and any two solutions differ by a constant. Thus

\vspace{1.5mm}
\hspace{2.7cm} $H^0([\alpha], ~\F)\simeq \C$, 

\hspace{2.7cm} $H^1([\alpha], ~\F) = 0$.
\end{proposition}
\noindent Proof: For $g \in \F$, define $f$ by 
\vspace{1mm}

(i) $f(0) = 0,$

(iI) $f(x) = \sum_{i=0}^{x-1}g(i)$, for $x \geqslant 1$,

(iii) $f(x) = - \sum_{i=x}^{-1} g(i)$, for $x \leqslant -1$.

\vspace{1mm}
Then $\alpha(f) - f = g$. Thus every $g \in \F$ is summable. 

\vspace{1.5mm}
The constants of $(\F, ~\alpha)$ are all the constant functions on $\Z$, hence 
$H^0([\alpha], ~\F)\simeq \C$.

If a 1-cochain $\theta: [\alpha] \rightarrow \F$ is a cocycle, then $\theta(\alpha)$ determines the values $\theta(\alpha^i)$ for all $i \in \Z$, just as in Proposition 7.2. As $\theta(\alpha)$ is summable in $(\F, ~\alpha)$, let it equal $\alpha(f) - f$, for some $f \in \F$. It follows, again as in Proposition 7.2, that $\theta(\alpha^i) = \alpha^i(f) - f$, for all $i$, and hence that $\theta = \delta(f)$. Thus every 1-cocycle is a coboundary and $H^1([\alpha], ~\F) = 0$. \hspace*{\fill}$\square$\\

Propositions 7.2 and 7.3 together constitute a `Poincar\'{e} duality' result for the classical ISP.


\section{acknowledgement} I am grateful to Shaoshi Chen for bringing to my attention his important article \cite{chen}. I am also grateful to Ananth Shankar for  many useful conversations.

\end{document}